\documentclass[conference,a4paper]{IEEEtran}

\usepackage{amsfonts}
\usepackage{cite}
\usepackage{graphicx}
\usepackage{amssymb}
\usepackage{subfigure}
\usepackage{array}
\usepackage{color}
\usepackage{amsmath}
\usepackage{algorithmic}
\usepackage{algorithm}
\usepackage{cases}
\usepackage{mathrsfs}
\usepackage{epsfig}
\usepackage{psfrag}
\usepackage{url}
\usepackage{setspace}
\usepackage{stfloats}
\usepackage{enumerate}
\usepackage{enumitem}
\allowdisplaybreaks[4]

\usepackage{etoolbox}
\makeatletter
\patchcmd{\@makecaption}
  {\scshape}
  {}
  {}
  {}
\makeatletter
\patchcmd{\@makecaption}
  {\\}
  {.\ }
  {}
  {}
\makeatother

\newtheorem{Thm}{Theorem}
\newtheorem{Lem}{Lemma}

\newtheorem{Prob}{Problem}

\newtheorem{Asump}{Assumption}

\IEEEoverridecommandlockouts

\begin{document}

\title{Stochastic Successive Convex Approximation for General Stochastic Optimization Problems with Applications in Wireless Communications}

\author{\IEEEauthorblockN{Chencheng Ye}\IEEEauthorblockA{Shanghai Jiao Tong University}
\and \IEEEauthorblockN{Ying Cui} \IEEEauthorblockA{Shanghai Jiao Tong University}}


\maketitle

\begin{abstract}
One key challenge for solving a general stochastic optimization problem with expectations in the objective and constraint functions using ordinary stochastic iterative methods lies in the infeasibility issue caused by the randomness over iterates. This letter aims to address this main challenge.
First, we obtain an equivalent stochastic optimization problem which is to minimize the weighted sum of the original objective and the penalty for violating the original constraints. Then, we propose a stochastic successive convex approximation (SSCA) method to obtain a stationary point of the original stochastic optimization problem.
Using similar techniques, we propose a parallel SSCA method to obtain a stationary point of a special case of the general stochastic optimization problem which has decoupled constraint functions.
We also provide application examples of the proposed methods in power control for interference networks.
The proposed SSCA and parallel SSCA methods achieve empirically higher convergence rates and lower computational complexities than existing ones, benefiting from the elegant way of balancing the objective minimization and constraint satisfaction over random iterates.
\end{abstract}


\section{Introduction}
Stochastic optimization problems refer to optimization problems which involve random variables. They are of broad interest, with applications arising in wireless communications, business analytics, manufacturing, finance, etc. 
In general, stochastic optimization problems with expectations in possibly nonconvex objective functions can be classified into three categories, namely, unconstrained stochastic problems, stochastic optimization problems with deterministic constraints, and stochastic optimization problems with expectations in constraint functions. Without loss of generality, in the following, we restrict our attention to stochastic minimization problems.

In~\cite{Bertsekas}, a stochastic gradient method is proposed to obtain a stationary point of an unconstrained stochastic optimization problem.
In~\cite{Ruszczynski,Mairal,Yang}, the stochastic gradient averaging method~\cite{Ruszczynski}, stochastic majorization-minimization (MM) method~\cite{Mairal} and stochastic successive convex approximation (SSCA) method~\cite{Yang} are proposed to obtain a stationary point of a stochastic optimization problem with deterministic convex constraints. 
Stochastic optimization problems with expectations in constraint functions are more challenging, as the stochastic nature of the constraint functions may cause infeasibility at each iteration of an ordinary stochastic iterative method.
In \cite{Liu}, an SSCA method is proposed to directly tackle a general stochastic optimization problem with expectations in the constraint functions, for the first time.
Specifically, at each iteration, an approximate convex problem is solved to minimize the objective; if it is infeasible, another approximate convex problem is then solved to minimize the penalty for violating the constraints.
Using similar techniques, a parallel SSCA method is proposed for a special case of the general stochastic optimization problem with decoupled constraint functions.
Leveraging two types of approximate problems with different goals at each iteration can deal with the infeasibility issue, but may lead to decrease of convergence rate and increment of computational complexity.

In this letter, we shall address the above issue. As in~\cite{Liu}, we consider a general stochastic optimization problem with expectations in the constraint functions. 
First, we obtain an equivalent stochastic optimization problem whose objective function is the weighted sum of the original objective and the penalty for violating the original constraints. Then, we propose an SSCA method that involves solving an approximate convex optimization problem which is always feasible at each iteration. 
Moreover, we show that the proposed SSCA method converges to a stationary point of the equivalent stochastic optimization problem, which is also a stationary point of the original stochastic optimization problem under certain conditions.
Using similar techniques, we propose a parallel SSCA method to obtain a stationary point of a special case of the aforementioned general stochastic optimization problem which has decoupled constraint functions.
As application examples, we consider the optimal power allocation to maximize the ergodic sum-rate under the coupled and decoupled individual ergodic rate constraints, respectively, and illustrate how to apply the proposed SSCA and parallel SSCA methods to obtain their respective stationary points.
Numerical results show that the proposed SSCA and parallel SSCA methods have higher convergence rates and lower computational complexities than those in~\cite{Liu}.
The substantial gains derive from the effective balance of the minimization of the original objective and the satisfaction of the original constraints over random iterates.

\section{General Stochastic Optimization}\label{Sec:gso}
In this section, we consider a general stochastic optimization problem with expectations in both the objective and constraint functions that are possibly nonconvex.
\begin{Prob}[General Stochastic Optimization Problem]\label{prob:gen}
\begin{align}
\min_{\mathbf{x}}\quad &f_0(\mathbf x) \triangleq\mathbb{E}\left[g_0(\mathbf x,\boldsymbol\xi)\right]\nonumber\\
\text{s.t.}  \quad &f_i(\mathbf x) \triangleq\mathbb{E}\left[g_i(\mathbf x,\boldsymbol\xi)\right]\leq0,\quad i=1,\dots,m,\label{eqn:f0}\\
&\mathbf{x}\in\mathcal{X},\label{eqn:xset}
\end{align}
where $\mathbf x\triangleq(x_1,\dots,x_n)$ is the optimization variable, $\mathcal{X}\subseteq\mathbb{R}^{n}$, $\boldsymbol\xi$ is a random vector defined on the probability space $(\Omega,\mathcal{F},\mathbb{P})$ with $\Omega$ being the sample space, $\mathcal{F}$ being the $\sigma$-algebra generated by subsets of $\Omega$, and $\mathbb{P}$ being a probability measure defined on $\mathcal{F}$, and functions $g_i: \mathcal{X}\times\Omega\to\mathbb{R}$, $i=0,\dots,m$ are possibly nonconvex.
\end{Prob}

\begin{Asump}[Assumptions on Problem~\ref{prob:gen}~\cite{Yang,Liu}]\label{asump:gi}
\begin{enumerate}
\item  $\mathcal{X}$ is compact and convex;
\item For any given $\boldsymbol\xi$, each $g_i(\mathbf x,\boldsymbol\xi)$ is continuously differentiable on $\mathcal{X}$, and its gradient is Lipschitz continuous.
\end{enumerate}
\end{Asump}

Problem~\ref{prob:gen} is very challenging, and is not well studied. First, motivated by the feasible point pursuit method in \cite{Mehanna}, we transform Problem~\ref{prob:gen} to the following stochastic optimization problem whose objective function is the weighted sum of the original objective and the penalty for violating the original constraints.
\begin{Prob}[Equivalent Problem of Problem~\ref{prob:gen}]\label{prob:pen}
\vspace{-3mm}
\begin{align}
\min_{\mathbf{x},\mathbf s}\quad &f_0(\mathbf x) +\rho\sum_{i=1}^m s_i\nonumber\\
\text{s.t.}  \quad &\eqref{eqn:xset},\nonumber\\
&f_i(\mathbf x) \leq s_i,\quad i=1,\dots,m,\label{eqn:fs}\\
&s_i\geq0,\quad i=1,\dots,m,\label{eqn:s0}
\end{align}
where $\mathbf{s}\triangleq(s_i)_{i=1,\dots,m}$ are slack variables and $\rho>0$ is a penalty parameter that trades off the original objective function and the slack penalty term.
\end{Prob}

Note that Problem~\ref{prob:pen} is always feasible. The relationship between Problem~\ref{prob:gen} and Problem~\ref{prob:pen} is summarized below.
\begin{Lem}[Equivalence between Problem~\ref{prob:gen} and Problem~\ref{prob:pen}]\label{lem:eq}
If Problem~\ref{prob:gen} is feasible and Assumption~\ref{asump:gi} is satisfied, then there exists $\rho_0\geq0$ such that for all $\rho\geq\rho_0$, Problem~\ref{prob:pen} and Problem~\ref{prob:gen} have the same optimal value.
\end{Lem}
\begin{IEEEproof}
As $g_0(\mathbf x,\boldsymbol\xi)$ is continuous and $\mathcal{X}$ is compact, $g_0(\mathbf x,\boldsymbol\xi)$ is bounded on $\mathcal{X}$. Thus, the optimal value of Problem~\ref{prob:gen} is bounded if Problem~\ref{prob:gen} is feasible. Therefore, by~\cite{Phan}, we know that there exists $\rho_0\geq0$ such that for all $\rho\geq\rho_0$, Problem~\ref{prob:pen} and Problem~\ref{prob:gen} have the same optimal value.
\end{IEEEproof}

Based on Lemma~\ref{lem:eq}, we now focus on solving Problem~\ref{prob:pen}. Like Problem~\ref{prob:gen}, Problem~\ref{prob:pen} is a stochastic optimization problem with possibly nonconvex objective and constraint functions. 
In the following, we propose an effective SSCA method to obtain a stationary point of Problem~\ref{prob:pen} using the SSCA technique~\cite{Yang}.
Later, we shall show that under certain conditions, a stationary point of Problem~\ref{prob:pen} is also a stationary point of Problem~\ref{prob:gen}.

Specifically, at iteration $t$, we solve the following approximate convex optimization problem of Problem~\ref{prob:pen}.
\begin{Prob}[Approximate Convex Optimization Problem in $t$-th Iteration]\label{prob:app}
\vspace{-3mm}
\begin{align}
\min_{\mathbf{x},\mathbf s}\quad &\bar{f}^t_0(\mathbf x)+\rho\sum_{i=1}^m s_i\nonumber\\
	\text{s.t.}  \quad &\eqref{eqn:xset},\eqref{eqn:s0},\nonumber\\
&\bar{f}^t_i(\mathbf x) \leq s_i,\quad i=1,\dots,m,
\end{align}
where $\bar f^t_i(\mathbf x),i=0,\dots,m$ are convex surrogate functions of $f_i(\mathbf x),i=0,\dots,m$. Let $(\bar{\mathbf{x}}^{t},\mathbf{s}^t)$ denote an optimal solution of Problem~\ref{prob:app}.
\end{Prob}

Problem~\ref{prob:app} is a convex optimization problem which is always feasible and can be solved with conventional convex optimization techniques.
Given $\bar{\mathbf{x}}^{t}$, we update $\mathbf x^t$ according to:
\begin{align}
&\mathbf{x}^{t}=(1-\gamma^{t})\mathbf{x}^{t-1}+\gamma^{t}\bar{\mathbf{x}}^{t},\ t=1,2,\dots,\label{eqn:updateTm}
\end{align}
where $\gamma^{t}$ is a positive diminishing stepsize satisfying:
\begin{align}
&\gamma^{t}=0,\ \lim_{t\to\infty}\gamma^{t}=0,\  \sum_{t=1}^\infty\gamma^{t}=\infty,\ \sum_{t=1}^\infty\left(\gamma^{t}\right)^2<\infty. \label{eqn:gamma}
\end{align}

The details are summarized in Alg.~\ref{alg:pssca}. 
To ensure the convergence of Alg.~\ref{alg:pssca}, the surrogate functions $\bar f^t_i(\cdot),i=0,\dots,m$ should satisfy the following assumptions.
\begin{Asump}[Assumptions on $\bar{f}_i^t(\mathbf{\cdot})$~\cite{Yang,Liu}]\label{asump:fi}
\begin{enumerate}
\item Each $\bar{f}_i^t(\mathbf{x})$ is uniformly strongly convex on $\mathcal{X}$;
\item Each $\bar{f}_i^t(\mathbf{x})$ is Lipschitz continuous on $\mathcal{X}$, and for any $\mathbf{x}\in\mathcal{X}$, $\mathop{\lim\sup}_{t_1 t_2\to\infty}\bar{f}_i^{t_1}(\mathbf{x})-\bar{f}_i^{t_2}(\mathbf{x})\leq B\Vert\mathbf{x}^{t_1}-\mathbf{x}^{t_2}\Vert$, for some constant $B\geq0$;
\item Each $\{\nabla^{2}_{\mathbf{x}}\bar{f}_i^t(\mathbf{x}):t=0,1,\dots\}$ is uniformly bounded;
\item $\lim_{t\to\infty}\vert\bar{f}_i^t(\mathbf{x}^t)-{f}_i^t(\mathbf{x}^t)\vert=0$ and $\lim_{t\to\infty}\Vert\nabla\bar{f}_i^t(\mathbf{x}^t)-\nabla{f}_i^t(\mathbf{x}^t)\Vert=0$.
\end{enumerate}
\end{Asump}

A common example of surrogate functions is~\cite{Yang,Liu}:
\begin{align}
\bar{f}^t_i(\mathbf{x})=&(1-\omega^{t})\bar{f}^{t-1}_i(\mathbf{x})+\omega^{t}\hat{g}_i(\mathbf{x},\mathbf{x}^{t-1},\boldsymbol\xi^t),\nonumber\\
&i=0,\dots,m, \quad t=1,2,\dots,\label{eqn:updatefi}
\end{align}
where $\bar{f}^0_i(\mathbf{x})=0$ for all $\mathbf{x}\in\mathcal{X}$,
$\omega^t$ is a positive diminishing stepsize satisfying:
\vspace{-3mm}
\begin{align}
&\omega^t>0,\quad \lim_{t\to\infty}\omega^t=0,\quad \sum_{t=1}^\infty\omega^t=\infty,\nonumber\\
&\sum_{t=1}^\infty\left(\omega^t\right)^2<\infty,\quad \lim_{t\to\infty}\frac{\gamma^{t}}{\omega^{t}}=0, \label{eqn:omega}
\end{align}
and $\hat{g}_i(\mathbf{x},\mathbf{x}^t,\boldsymbol\xi^t)$ is a convex approximation of ${g}_i(\mathbf{x},\boldsymbol\xi^{t-1})$
around $\mathbf x^{t-1}$ satisfying:
$\hat{g}_i(\mathbf{x},\mathbf{x},\boldsymbol\xi)={g}_i(\mathbf{x},\boldsymbol\xi)$ and $\nabla\hat{g}_i(\mathbf{x},\mathbf{x},\boldsymbol\xi)=\nabla{g}_i(\mathbf{x},\boldsymbol\xi)$, for all $\mathbf{x}\in\mathcal{X}$ and $\boldsymbol\xi\in\Omega$;
$\hat{g}_i(\mathbf{x},\mathbf{y},\boldsymbol\xi)$ is strongly convex in $\mathbf x$ for all $\mathbf{y}\in\mathcal{X}$ and $\boldsymbol\xi\in\Omega$;
$\hat{g}_i(\mathbf{x},\mathbf{y},\boldsymbol\xi)$ is Lipschitz continuous in both $\mathbf x$ and $\mathbf y$ for all $\boldsymbol\xi\in\Omega$.
It has been shown in \cite[Proposition 1]{Liu} that if the stepsizes $\{\gamma^{t}\}$ and $\{\omega^{t}\}$ satisfy \eqref{eqn:gamma} and \eqref{eqn:omega}, respectively, then the surrogate functions given by \eqref{eqn:updatefi} satisfy Assumption~\ref{asump:fi}.

Finally, we show the convergence of Alg.~\ref{alg:pssca}. 
\begin{Thm}[Convergence of Alg.~\ref{alg:pssca}]\label{thm:conv-rand}
Suppose Assumption~\ref{asump:gi} and Assumption~\ref{asump:fi} are satisfied. Then $\{(\mathbf{x}^{t},\mathbf{s}^{t})\}$ generated by Alg.~\ref{alg:pssca} has a limit point, denote by $(\mathbf x^*,\mathbf s^*)$, and the following statements hold:
\begin{enumerate}
\item $(\mathbf x^*,\mathbf s^*)$ is a stationary point of Problem~\ref{prob:pen};
\item If $\mathbf s^*=\mathbf0$, then $\mathbf x^*$ is a stationary point of Problem~\ref{prob:gen}.
\end{enumerate}
\end{Thm}
\begin{IEEEproof}
By Assumption~\ref{asump:gi}.1, we know that $\mathbf x^t$ is bounded. As $(\bar{\mathbf{x}}^{t},\mathbf{s}^t)$ is an optimal solution of Problem~\ref{prob:app}, it can be easily shown that $s^t_i=\bar{f}^t_i(\bar{\mathbf x}^t)$, $i=1,\dots,m$. 
By Assumption~1.1 and Assumption~2.2, we know that $\bar{f}^t_i(\bar{\mathbf x}^t)$ is bounded, which implies that $\mathbf{s}^t$ is bounded. Note that Assumption~1.2, Assumption~2 and Assumption~3 in~\cite{Liu} readily follow Assumption 1.2 and Assumption~2 in this letter, and Assumption~1.1 in~\cite{Liu} is used to prove the boundedness of $\bar{f}^t_i(\bar{\mathbf x}^t)+\mathbf{s}^t$. Therefore, following the proof of \cite[Theorem 1]{Liu}, we can show the first statement.
In addition, it can be easily shown that when $\mathbf s=\mathbf 0$, the KKT conditions of Problem~\ref{prob:pen} imply those of Problem~\ref{prob:gen}.
Therefore, we can show the second statement.
\end{IEEEproof}

Note that we can run Alg.~\ref{alg:pssca} multiple times, each with a random initial point $\mathbf{x}^{0}\in\mathcal{X}$, until a stationary point $(\mathbf x^*,\mathbf s^*)$ of Problem~\ref{prob:pen} with $\mathbf s^*=\mathbf 0$, i.e., a stationary point $\mathbf x^*$ of Problem~\ref{prob:gen}, is obtained.

\begin{algorithm}[t]
    \caption{SSCA}
\begin{small}
        \begin{algorithmic}[1]
           \STATE \textbf{initialization}: Set $t=1$, and choose any $\mathbf{x}^{0}\in\mathcal{X}$.\\
           \STATE \textbf{repeat}
           \STATE \quad Obtain $(\bar{\mathbf{x}}^{t},\mathbf{s}^{t})$ by solving Problem~\ref{prob:app} with conventional convex optimization techniques, and update $\mathbf{x}^{t}$ according to \eqref{eqn:updateTm}.
           \STATE\quad Set $t=t+1$.
           \STATE \textbf{until} Some convergence criteria is met.
    \end{algorithmic}\label{alg:pssca}
    \end{small}
    \vspace{-0.1cm}
\end{algorithm}

\section{Stochastic Optimization with Decoupled Constraints}\label{Sec:pd}
In this section, we consider a special case of the general stochastic optimization problem, which has decoupled constraint functions involving expectations.
\begin{Prob}[Stochastic Optimization Problem with Decoupled Constraints]\label{prob:dec-gen}
\begin{align}
\min_{\mathbf x}\  &f_0(\mathbf x) \triangleq\mathbb{E}\left[g_0(\mathbf x,\boldsymbol\xi)\right]\nonumber\\
	\text{s.t.}  \  &f_{k,i}(\mathbf x_k) \triangleq\mathbb{E}\left[g_{k,i}(\mathbf x_k,\boldsymbol\xi)\right]\leq0,k=1,...,K,i=1,...,m_k,\\
	&\mathbf x_k\in\mathcal{X}_k,\ k=1,...,K,\label{eqn:xkset}
\end{align}
where the optimization variable can be partitioned into $K$ blocks, i.e., $\mathbf x\triangleq(\mathbf x_k)^K_{k=1}$, with $\mathbf x_k$ being the variable for the $k$-th block,  $\mathcal{X}_k\subseteq\mathbb{R}^{n_k}$, $k=1,...,K$,
$\boldsymbol\xi$ is a random vector, and functions $g_{0}: \mathcal{X}\times\Omega\to\mathbb{R}$ with $\mathcal{X}\triangleq\mathcal{X}_1\times\cdots\times\mathcal{X}_K\subseteq\mathbb{R}^{n}$ and $g_{k,i}: \mathcal{X}_k\times\Omega\to\mathbb{R}$, $k=1,...,K,i=1,...,m_k$ are possibly nonconvex. 
\end{Prob}

\begin{Asump}[Assumptions on Problem~\ref{prob:dec-gen}~\cite{Liu}]\label{asump:gik}
\begin{enumerate}
\item  $\mathcal{X}$ is compact and convex;
\item For any given $\boldsymbol\xi$, $g_{0}(\mathbf x,\boldsymbol\xi)$ and each $g_{k,i}(\mathbf x_k,\boldsymbol\xi)$ are continuously differentiable on $\mathcal{X}$ and $\mathcal{X}_k$, respectively, and their gradients are Lipschitz continuous.
\end{enumerate}
\end{Asump}

Note that the constraints of Problem~\ref{prob:dec-gen} can be separated into $K$ groups with the $k$-th group of constraints depending on the $k$-th block $\mathbf x_k$.
Similarly, we transform Problem~\ref{prob:dec-gen} to the following stochastic optimization problem.
\begin{Prob}[Equivalent Problem of Problem~\ref{prob:dec-gen}]\label{prob:dec-pen}
\vspace{-3mm}
\begin{align}
\min_{\mathbf x,\mathbf s}\quad &f_0(\mathbf x)+\rho\sum_{k=1}^K\sum_{i=1}^{m_k} s_{k,i}\nonumber\\
	\text{s.t.}  \quad &\eqref{eqn:xkset},\nonumber\\
	&f_{k,i}(\mathbf x_k) \leq s_{k,i},\ k=1,\dots,K,\  i=1,\dots,m_k,\\
	&s_{k,i}\geq0, \ k=1,\dots,K,\  i=1,\dots,m_k,\label{eqn:sk0}
\end{align}
where $\mathbf{s}\triangleq(s_{k,i})_{i=1,\dots,m_k,k=1,\dots,K}$ are slack variables and $\rho>0$ is the penalty parameter.
\end{Prob}

Note that Problem~\ref{prob:dec-pen} is always feasible. The relationship between Problem~\ref{prob:dec-gen} and Problem~\ref{prob:dec-pen} is summarized below.
\begin{Lem}[Equivalence between Problem~\ref{prob:dec-gen} and Problem~\ref{prob:dec-pen}]\label{lem:eq2}
If Problem~\ref{prob:dec-gen} is feasible and Assumption~\ref{asump:gik} is satisfied, then there exists $\rho_0\geq0$ such that for all $\rho\geq\rho_0$, Problem~\ref{prob:dec-pen} and Problem~\ref{prob:dec-gen} have the same optimal value.
\end{Lem}
\begin{IEEEproof}
The proof is similar to that of Lemma~\ref{lem:eq}, and is omitted due to page limitation.
\end{IEEEproof}

Based on Lemma~\ref{lem:eq2}, we now focus on solving Problem~\ref{prob:dec-pen}.
In the following, we propose an effective parallel SSCA method to obtain a stationary point of Problem~\ref{prob:dec-pen} using the parallel SSCA technique~\cite{Yang}.
Similarly, we shall show that under certain conditions, a stationary point of Problem~\ref{prob:dec-pen} is also a stationary point of Problem~\ref{prob:dec-gen}.

Specifically, at iteration $t$, we solve the following $K$ approximate convex optimization problems of Problem~\ref{prob:dec-pen}, one for each block.

\begin{Prob}[Approximate Convex Optimization Problem for $k$-th Block in $t$-th Iteration]\label{prob:dec-app}
\vspace{-3mm}
\begin{align}
\min_{\mathbf x_k,\mathbf s_k}\quad &\bar{f}^t_{k,0}(\mathbf x_k)+\rho\sum_{i=1}^{m_k} s_{k,i}\nonumber\\
\text{s.t.}  \quad &\bar{f}^t_{k,i}(\mathbf x_k) \leq s_{k,i}, i=1,\dots,m_k,\\
&s_{k,i}\geq0, i=1,\dots,m_k,\\
&\mathbf x_k\in\mathcal{X}_k,
\end{align}
\end{Prob}
where $\mathbf s_k\triangleq(s_{k,i})_{i=1,\dots,m_k}$, $\bar f^t_{k,0}(\mathbf x)$, $\bar f^t_{k,i}(\mathbf x), i=1,\dots,m_k$ are convex surrogate functions of $f_{0}(\mathbf x)$, $f_{k,i}(\mathbf x), i=1,\dots,m_k$.\footnote{
Note that the surrogate objective function of the parallel SSCA method in this letter is more general than that in~\cite{Liu}, and can exploit block-wise structures of the objective function.
}
Let $(\bar{\mathbf{x}}^{t}_k,\mathbf{s}^t_k)$ denote an optimal solution of Problem~\ref{prob:dec-app}.

The $K$ approximate problems can be solved in a distributed and parallel manner using conventional convex optimization techniques~\cite{Yang}.
Denote ${\mathbf{x}}^{t}\triangleq({\mathbf x}^t_k)^K_{k=1}$.
The details are summarized in Alg.~\ref{alg:parallel}.
To ensure the convergence of Alg.~\ref{alg:parallel}, the surrogate functions $\bar f^t_{k,i}(\cdot),i=0,\dots,m_k,k=1,\dots,K$ should satisfy the following assumptions.
\begin{Asump}[Assumptions on $\bar{f}_{k,i}^t(\cdot)$~\cite{Yang,Liu}]\label{asump:fik}
\begin{enumerate}
\item Each $\bar{f}_{k,i}^t(\mathbf{x}_k)$ is uniformly strongly convex on $\mathcal{X}_k$;
\item Each $\bar{f}_{k,i}^t(\mathbf{x}_k)$ is Lipschitz continuous on $\mathcal{X}_k$, and for any $\mathbf{x}_k\in\mathcal{X}_k$, $\mathop{\lim\sup}_{t_1 t_2\to\infty}\bar{f}_{k,i}^{t_1}(\mathbf{x}_k)-\bar{f}_{k,i}^{t_2}(\mathbf{x}_k)\leq B\Vert\mathbf{x}_k^{t_1}-\mathbf{x}_k^{t_2}\Vert$, for some constant $B\geq0$;
\item Each $\{\nabla^{2}_{\mathbf{x}_k}\bar{f}_{k,i}^t(\mathbf{x}_k):t=0,1,\dots\}$ is uniformly bounded;
\item $\lim_{t\to\infty}\vert\bar{f}_{k,i}^t(\mathbf{x}_k^t)-{f}_{k,i}^t(\mathbf{x}_k^t)\vert=0$ and $\lim_{t\to\infty}\vert\nabla\bar{f}_{k,i}^t(\mathbf{x}_k^t)-\nabla{f}_{k,i}^t(\mathbf{x}_k^t)\vert=0$.
\end{enumerate}
\end{Asump}

A common example of surrogate functions is given as follows~\cite{Yang,Liu}:
\begin{align}
&\bar{f}^t_{k,i}(\mathbf{x}_k)=(1-\omega^{t})\bar{f}^{t-1}_{k,i}(\mathbf{x}_k)+\omega^{t}\hat{g}_{k,i}(\mathbf{x}_k,\mathbf{x}^{t-1},\boldsymbol\xi^t),\nonumber\\
&i=0,\dots,m_k,\ k=1,\dots,K, \ t=1,2,\dots,\label{eqn:updatefi}
\end{align}
where $\bar{f}^0_{k,i}(\mathbf{x}_k)=0$ for all $\mathbf{x}_k\in\mathcal{X}_k$,
$\omega^t$ is a positive diminishing stepsize satisfying \eqref{eqn:omega},
and $\hat{g}_{k,i}(\mathbf{x}_k,\mathbf{x}^t,\boldsymbol\xi^{t-1})$ is a convex approximation of ${g}_{k,i}(\mathbf{x}_k,\mathbf{x}^t_{-k},\boldsymbol\xi^t)$ around $\mathbf x_k^{t-1}$ satisfying:
$\hat{g}_{k,i}(\mathbf{x}_k,\mathbf{x},\boldsymbol\xi)={g}_{k,i}(\mathbf{x},\boldsymbol\xi)$ and $\nabla_{\mathbf x_k}\hat{g}_{k,i}(\mathbf{x}_k,\mathbf{x},\boldsymbol\xi)=\nabla_{\mathbf x_k}{g}_{k,i}(\mathbf{x},\boldsymbol\xi)$, for all $\mathbf{x}\in\mathcal{X}$ and $\boldsymbol\xi\in\Omega$;
$\hat{g}_{k,i}(\mathbf{x}_k,\mathbf{y},\boldsymbol\xi)$ is strongly convex in $\mathbf x_k$ for all $\mathbf{y}\in\mathcal{X}$ and $\boldsymbol\xi\in\Omega$;
$\hat{g}_{k,i}(\mathbf{x}_k,\mathbf{y},\boldsymbol\xi)$ is Lipschitz continuous in both $\mathbf x_k$ and $\mathbf y$ for all $\boldsymbol\xi\in\Omega$.
Similarly, by~\cite[Proposition 1]{Liu}, the surrogate functions given by \eqref{eqn:updatefi} satisfy Assumption~\ref{asump:fik}.

Finally, we show the convergence of Alg.~\ref{alg:parallel}. 
\begin{Thm}[Convergence of Alg.~\ref{alg:parallel}]\label{thm:conv-rand2} 
Suppose Assumption~\ref{asump:gik} and Assumption~\ref{asump:fik} are satisfied. Then $\{(\mathbf{x}^{t},\mathbf{s}^{t})\}$ generated by Alg.~\ref{alg:parallel} has a limit point, denote by $(\mathbf x^\star,\mathbf s^\star)$, and the following statements hold:
\begin{enumerate}
\item $(\mathbf x^\star,\mathbf s^\star)$ is a stationary point of Problem~\ref{prob:dec-pen};
\item If $\mathbf s^\star=\mathbf 0$, then $\mathbf x^\star$ is a stationary point of Problem~\ref{prob:dec-gen}.
\end{enumerate}
\end{Thm}

\begin{IEEEproof}
Similarly to the proof of Theorem~\ref{alg:pssca}, we can show that $\mathbf{s}^t$ is bounded. Note that Assumption~b, Assumption~c in~\cite{Yang} readily follow Assumption 1.2 and Assumption~2 in this letter, and Assumption~a in~\cite{Yang} is used to prove the boundedness of $\bar{f}^t_i(\bar{\mathbf x}^t)+\mathbf{s}^t$, where $\bar{\mathbf{x}}^{t}\triangleq(\bar{\mathbf x}^t_k)^K_{k=1}$. Thus, following the proof of \cite[Theorem~1]{Yang}, we can show  Lemma~4 in~\cite{Liu}. Then, following the proof of \cite[Theorem~1]{Liu}, we can show
the KKT conditions of Problem~\ref{prob:dec-pen} hold. Therefore, we can show the first statement.
Similarly, when $\mathbf s=\mathbf 0$, the KKT conditions of Problem~\ref{prob:dec-pen} imply those of Problem~\ref{prob:dec-gen}. Therefore, we can show the second statement.
%
\end{IEEEproof}
Similarly, we can run Alg.~\ref{alg:parallel} multiple times, each with a random initial point $\mathbf{x}^{0}\in\mathcal{X}$, until a stationary point of Problem~\ref{prob:dec-gen} is obtained.

\begin{algorithm}[t]
    \caption{Parallel SSCA}
\begin{small}
        \begin{algorithmic}[1]
           \STATE \textbf{initialization}: Set $t=1$, and choose any $\mathbf{x}^{0}\in\mathcal{X}$.\\
           \STATE \textbf{repeat}
           \STATE \quad Obtain $(\bar{\mathbf{x}}_k^{t},\mathbf{s}_k^{t})$ by solving Problem~\ref{prob:dec-app} with conventional convex optimization techniques, for $k=1,\dots,K$, and update $\mathbf{x}^{t}$ according to \eqref{eqn:updateTm}.
           \STATE\quad Set $t=t+1$.
           \STATE \textbf{until} Some convergence criteria is met.
    \end{algorithmic}\label{alg:parallel}
    \end{small}
    \vspace{-0.1cm}
\end{algorithm}

\section{Application examples in interference networks}\label{sec:example}
Consider a $K$-pair frequency-selective interference channel. 
Each pair includes one single-antenna transmitter and one single-antenna receiver. Let $H_{kj}$ denote the random coefficient of the channel between the $k$-th transmitter and the $j$-th receiver. Suppose $H_{kj}$, $k,j=1,\dots,K$, are independent and identically distributed according to $\mathcal{CN}(0,\delta_{kj})$, $k,j=1,\dots,K$.
Let $p_k$ denote the transmit power for the $k$-th transmitter, where
\begin{align}
0\leq p_k\leq P_k,\quad k=1,\dots,K.\label{eqn:power}
\end{align}
Here, $P_k$ represents the power limit for the $k$-th transmitter. Denote $\mathbf{p}\triangleq(p_k)_{k=1}^K$. The ergodic rate of the $k$-th pair is given by
$r_k(\mathbf{p})=\mathbb{E}\left[\log\left(1+\frac{\vert{H_{kk}}\vert^2 p_k}{\sum_{j\neq k}\vert{H_{kj}}\vert^2 p_j+\sigma_k^2}\right)\right]$,
where $\sigma_k$ denotes the variance of the additive complex Gaussian noise at the $k$-th receiver. 
The ergodic sum-rate of the $K$ pairs is given by
$r_0(\mathbf{p})=\sum_{k=1}^K r_k(\mathbf{p})$.
The ergodic rate of the $k$-th pair satisfies:
\begin{align}
r_k(\mathbf{p})\geq R_k,\quad k=1,\dots,K,\label{eqn:ratec1}
\end{align}
where $R_k$ represents the rate requirement for the $k$-th pair. 
By~\eqref{eqn:power}, 
$r_k(\mathbf p)\geq\mathbb{E}\left[\log(1+\frac{\vert{H_{kk}}\vert^2 p_k}{\sum_{j\neq k}\vert{H_{kj}}\vert^2 P_j+\sigma_k^2})\right]\triangleq r_{lb,k}(p_k).$
Thus, a stronger and decoupled version of \eqref{eqn:ratec1} is given by:
\begin{align}
r_{lb,k}(\mathbf{p})\geq R_k,\quad k=1,\dots,K. \label{eqn:ratec2}
\end{align}

We would like to optimize the transmit power $\mathbf p$ to maximize the ergodic sum-rate $r_0(\mathbf p)$, subject to the power constraints in \eqref{eqn:power} as well as the coupled and decoupled individual ergodic rate constraints in \eqref{eqn:ratec1} and \eqref{eqn:ratec2}, respectively.
\begin{Prob}[Ergodic Sum-Rate Maximization with Coupled Constraints]\label{prob:sumrate-couple}
\begin{align}
\max_{\mathbf{p}}\quad &r_0(\mathbf{p})\nonumber\\
\text{s.t.}  \quad &\eqref{eqn:power},\eqref{eqn:ratec1}.\nonumber
\end{align}
\end{Prob}

\begin{figure*}
\begin{small}
\begin{align}
&\hat{g}_{0}(\mathbf{p},\mathbf{p}^{t-1},\mathbf{H}^t)=\sum_{k=1}^K\Big(\log({\sum_{j=1}^K\vert{H_{kj}}\vert^2 p_j+\sigma_k})-\log({\!\!\!\sum_{{j=1,j\neq k}}^K\!\!\!\vert{H_{kj}}\vert^2 p^t_j+\sigma_k})-\frac{\sum_{{j=1,j\neq k}}^K\vert{H_{kj}}\vert^2(p_j-p^t_j)}{\sum_{{l=1,l\neq k}}^K\vert{H_{kl}}\vert^2 p^t_l+\sigma_k}\Big) \label{eqn:exg0}\\
&\hat{g}_{k}(\mathbf{p},\mathbf{p}^{t-1},\mathbf{H}^t)=R_k-\log({\sum_{j=1}^K\vert{H_{kj}}\vert^2 p_j+\sigma_k})+\log({\!\!\!\sum_{{j=1,j\neq k}}^K\!\!\!\vert{H_{kj}}\vert^2 p^t_j+\sigma_k})+\frac{\sum_{{j=1,j\neq k}}^K\vert{H_{kj}}\vert^2(p_j-p^t_j)}{\sum_{{l=1,l\neq k}}^K\vert{H_{kl}}\vert^2 p^t_l+\sigma_k},\ k=1,\dots,K \label{eqn:exgk}
\end{align}
\end{small}
\normalsize 
\vspace{-9mm}
\end{figure*}

\begin{Prob}[Ergodic Sum-Rate Maximization with Decoupled Constraints]\label{prob:sumrate-decouple}
\begin{align}
\max_{\mathbf{p}}\quad &r_0(\mathbf{p})\nonumber\\
\text{s.t.}  \quad &\eqref{eqn:power},\eqref{eqn:ratec2}.\nonumber
\end{align}
\end{Prob}

Problem~\ref{prob:sumrate-couple} is one instance of Problem~\ref{prob:gen}. We can choose $\hat{g}_{0}$ and $\hat{g}_{k}$, given by \eqref{eqn:exg0} and \eqref{eqn:exgk}, as shown at the top of this page, and obtain a stationary point of Problem~\ref{prob:sumrate-couple} using Alg.~\ref{alg:pssca}.
Problem~\ref{prob:sumrate-decouple} is one instance of Problem~\ref{prob:dec-gen}. We can choose $\hat{g}_{k,0}$ and $\hat{g}_{k,1}$, given by \eqref{eqn:exgk0} and \eqref{eqn:exgk1}, as shown at the top of this page,
\begin{figure*}
\begin{footnotesize}
\begin{align}
&\hat{g}_{k,0}(p_k,\mathbf{p}^{t-1},\mathbf{H}^t)=\sum_{m=1}^K\Big(\log({\vert{H_{mk}}\vert^2 p_k+\!\!\!\sum_{{j=1,j\neq k}}^K\!\!\!\vert{H_{mj}}\vert^2 p_j^t+\sigma_m})-\log({\!\!\!\sum_{{j=1,j\neq m}}^K\!\!\!\vert{H_{mj}}\vert^2 p^t_j+\sigma_m})-\frac{\vert{H_{mk}}\vert^2(p_k-p^t_k)}{\sum_{{l=1,l\neq m}}^K\vert{H_{ml}}\vert^2 p^t_l+\sigma_m}\Big) \label{eqn:exgk0}\\
&\hat{g}_{k,1}(\mathbf{p},\mathbf{p}^{t-1},\mathbf{H}^t)=R_k-\log({\vert{H_{kk}}\vert^2 p_k+\!\!\!\sum_{{j=1,j\neq k}}^K\!\!\! \vert{H_{kj}}\vert^2 P_j+\sigma_k})-\log({\!\!\!\sum_{{j=1,j\neq k}}^K\!\!\!\vert{H_{kj}}\vert^2 P_j+\sigma_k})-\frac{\vert{H_{kk}}\vert^2(p_k-p^t_k)}{\sum_{{l=1,l\neq k}}^K\vert{H_{kl}}\vert^2 P_l+\sigma_k},\ k=1,\dots,K \label{eqn:exgk1}
\end{align}
\end{footnotesize}
\normalsize \hrulefill
\vspace{-4mm}
\end{figure*}
and obtain a stationary point of Problem~\ref{prob:sumrate-decouple} using Alg.~\ref{alg:parallel}.
Problem~\ref{prob:sumrate-decouple} has a smaller optimal ergodic sum-rate than Problem~\ref{prob:sumrate-couple}, but yields a parallel SSCA method with faster convergence speed. Thus, Problem~\ref{prob:sumrate-decouple} has application when the network topology changes rapidly over time.

\section{Numerical Results}
In this section, we consider the application examples in Section \ref{sec:example}, and compare 
the proposed SSCA and parallel SSCA methods (i.e., Alg.~\ref{alg:pssca} and Alg.~\ref{alg:parallel}) with those in~\cite{Liu} through numerical experiments.
We set $K=5$ and $\rho=0.5$. For simplicity, we choose $P_k=100$, $\sigma^2_k=1$ and $R_k=1$ for all $k=1,\dots,K$. We choose $\delta_{kj}^2=1$ if $k=j$ and $\delta_{kj}^2=0.1$ otherwise. 
We independently generate 50 sample paths of random channel coefficients according to $\mathcal{CN}(0,\delta_{kj})$, $k,j=1,\dots,K$, and evaluate the average convergence rates and computing times.
We choose $p^{0}_k=P_k$, $k=1,\dots,K$ as the initial point of the four algorithms. For each generated sample path, Alg.~\ref{alg:pssca} and the SSCA method in~\cite{Liu} for solving Problem~\ref{prob:sumrate-couple} converge to the same stationary point, denoted by $\mathbf p^*$; Alg.~\ref{alg:parallel} and the parallel SSCA method in~\cite{Liu} for solving Problem~\ref{prob:sumrate-decouple} converge to the same stationary point, denoted by $\mathbf p^\star$.

\begin{figure}[t]
\begin{center}
\subfigure[\scriptsize{
Problem~\ref{prob:sumrate-couple}}\label{fig:couple}]
{\resizebox{5.2cm}{!}{\includegraphics{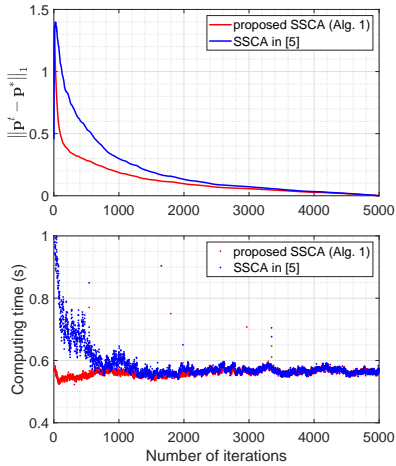}}}\quad
\subfigure[\scriptsize{
Problem~\ref{prob:sumrate-decouple}}\label{fig:decouple}]
{\resizebox{5.2cm}{!}{\includegraphics{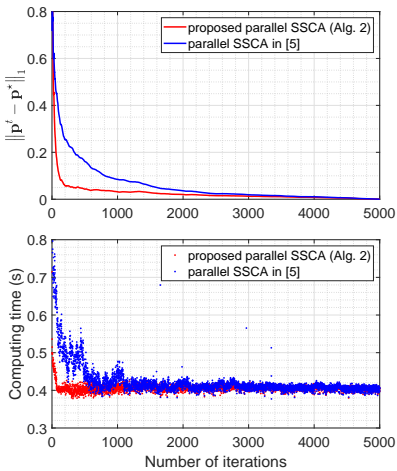}}}
\end{center}
\vspace{-4mm}
\caption{\small{Convergence rates and computing times.}}
\vspace{-2mm}
\label{fig:appl}
\end{figure}
 

Fig.~\ref{fig:appl} illustrates the convergence rates and computing times.
Table. 1 illustrates the numbers of iterations and total computing times when certain convergence criterion are satisfied.
From Fig.~\ref{fig:appl} and Table 1, we can see that the proposed SSCA and parallel SSCA methods have higher convergence rates and shorter computing times than those in~\cite{Liu}. The gains in convergence rate come from solving a single type of approximate convex problems over all iterates.
The gains in the computational complexity stem from solving a single optimization problem per iteration.
The substantial gains demonstrate the effectiveness for balancing the objective minimization and the constraint satisfaction over random iterates.

\begin{table}[t]
\centering
\caption{Numbers of iterations and total computing times at $\Vert{\mathbf p^t-\mathbf p^*}\Vert_1/\Vert{\mathbf p^*}\Vert_1=\Vert{\mathbf p^t-\mathbf p^\star}\Vert_1/\Vert{\mathbf p^\star}\Vert_1=0.02$.}
   \vspace{-0.3cm}
   \scriptsize{
\begin{tabular}{|c|c|c|c|c|}
\hline
&Alg.~1 &SSCA [5] &Alg.~2 &PSSCA [5]\\
\hline
Number of iterations &1956 &2390 &123 &725\\
\hline
Computing time (s)  &1084 &1455 &53  &378\\
\hline
\end{tabular}}
\label{parameter}
   \vspace{-0.5cm}
\end{table}

\section{Conclusion}
In this letter, we considered the general stochastic optimization problem with expectations in both the objective and constraint functions. 
We proposed a SSCA method and a parallel SSCA method to obtain stationary points of the general stochastic optimization problem and its special case with decoupled constraint functions, respectively. We provided application examples of the proposed methods and demonstrated the advantages of the proposed methods in terms of convergence rate and computational complexity.


\end{document}